\documentclass[onecolumn]{aastex6}

\usepackage{color,txfonts}
\usepackage{hyperref}
\usepackage{epstopdf}
\usepackage{graphicx}
\usepackage[FIGTOPCAP]{subfigure}
\usepackage{xcolor}

\begin{document}

\title{GW170817/GRB 170817A/AT2017gfo association: some implications for physics and astrophysics}

\author{Hao Wang$^{1,2}$, Fu-Wen Zhang$^{3}$, Yuan-Zhu Wang$^{1,2}$, Zhao-Qiang Shen$^{1,2}$, Yun-Feng Liang$^{1}$, Xiang Li$^{1}$, Neng-Hui Liao$^{1}$, Zhi-Ping Jin$^{1,4}$, Qiang Yuan$^{1,4}$,
 Yuan-Chuan Zou$^{5}$, Yi-Zhong Fan$^{1,4}$, and Da-Ming Wei$^{1,4}$}
\affil{
$^1$ {Key Laboratory of dark Matter and Space Astronomy, Purple Mountain Observatory, Chinese Academy of Science, Nanjing, 210008, China.}\\
$^2$ {University of Chinese Academy of Sciences, Yuquan Road 19, Beijing, 100049, China}\\
$^3$ {College of Science, Guilin University of Technology, Guilin 541004, China.}\\
$^4$ {School of Astronomy and Space Science, University of Science and Technology of China, Hefei, Anhui 230026, China.}\\
$^5$ {School of Physics, Huazhong University of Science and Technology, Wuhan 430074, China.}\\
}
\email{xiangli@pmo.ac.cn(XL) and yzfan@pmo.ac.cn (YZF)}

\begin{abstract}
On 17 August 2017, a gravitational wave event (GW170817) and an associated short gamma-ray burst (GRB 170817A) from a binary neutron star merger had been detected. The followup optical/infrared observations also identified the macronova/kilonova emission (AT2017gfo). In this work we discuss some implications of the remarkable GW170817/GRB 170817A/AT2017gfo association. We show that the $\sim 1.7$s time delay between the gravitational wave (GW) and GRB signals imposes very tight constraint on the superluminal movement of gravitational waves (i.e., the relative departure of GW velocity from the speed of light is $\leq 4.3\times 10^{-16}$) or the possible violation of weak equivalence principle (i.e., the difference of the gamma-ray and GW trajectories in the gravitational field of the galaxy and the local universe should be within a factor of $\sim 3.4\times 10^{-9}$). The so-called Dark Matter Emulators and a class of contender models for cosmic acceleration (``Covariant Galileon") are ruled out, too. The successful identification of Lanthanide elements in the macronova/kilonova spectrum also excludes the possibility that the progenitors of GRB 170817A are a binary strange star system.  The high neutron star merger rate (inferred from both the local sGRB data and the gravitational wave data) together with the significant ejected mass strongly suggest that such mergers are the prime sites of heavy r-process nucleosynthesis.
\end{abstract}

\keywords{}

\section{Introduction} \label{sec:intro}
The mergers of close compact object binaries are
promising gravitational-wave (GW) sources \citep{1977ApJ...215..311C}, as demonstrated by the successful detection of the mergers of three massive black hole binaries\citep{2016PhRvX...6d1015A,2016PhRvL.116f1102A,2017PhRvL.118v1101A}. Usually no electromagnetic counterparts are expected from the binary black hole mergers unless some pre-merger objects have massive accretion disks. Therefore the information that can be directly inferred are limited. For the compact object mergers involving at least one neutron star, the situation is dramatically different. These mergers are expected to launch ultra-relativistic ejecta and neutron-rich sub-relativistic outflows. The ultra-relativistic ejecta can give rise to sGRBs \citep{1989Natur.340..126E,1993ApJ...413L.101K,2004RvMP...76.1143P,2004IJMPA..19.2385Z} while the r-process nucleosynthesis takes place in the neutron-rich sub-relativistic outflows and then generates optical/infrared transients \citep[i.e., the so-called marconova or kilonova; see][]{1998ApJ...507L..59L,2005astro.ph.10256K,2010MNRAS.406.2650M,2013ApJ...774...25K,2013ApJ...775...18B,2013ApJ...775..113T,2017LRR....20....3M}. After the historical detection of the GW emission from binary black holes, people are looking forward to catching the neutron star mergers by the advanced LIGO/Virgo. The first electromagnetic counterpart of such GW events is widely believed to be macronova/kilonova since its emission is almost isotropic \citep{2012ApJ...746...48M} and moreover a few candidates have already been reported in GRB 130603B \citep{2013Natur.500..547T,2013ApJ...774L..23B}, GRB 060614 \citep{2015NatCo...6E7323Y, 2015ApJ...811L..22J} and GRB 050709 \citep{2016NatCo...712898J}. While the sGRBs are widely known to be beamed with a typical half-opening angle of $\sim 0.1$ rad, which will suppress the GRB/GW association very effectively. Therefore it is widely suspected that the first GRB/GW association will not be established in 2020s when the advanced LIGO/Virgo are running at their full sensitivity \citep{2015ApJ...809...53C, 2016ApJ...827L..16L}. Very recently it has been noticed that the GRB/GW association chance can be high up to $\sim 10\%$ since the neutron star merger events detectable for advanced LIGO/Virgo are very nearby and hence some off-beam events (if the ejecta are uniform) or the off-axis events (if the ejecta are structured) can still be detectable \citep{2017arXiv170807008J}. Even so, it is still less likely that the first neutron star merger GW event would be accompanied by a sGRB.

On 2017 August 17, the LIGO and Virgo detectors simultaneously detected a transient GW signal
that is consistent with the merger of a pair of neutron stars \citep{LVC2017}. Surprisingly, at 12:41:06.47 UT on 17 August 2017,
the Fermi Gamma-Ray Burst Monitor (GBM) triggered and located GRB 170817A \citep{von Kienlin2017}, which is just about 1.7
seconds after the GW signal and the location also overlaps with the GW event \citep[][]{Blackburn2017}.
The optical/infrared/ultraviolet followup observations \citep[e.g.][]{Coulter2017,Pian2017} found a bright unpolarized source \citep{Covino2017} and the high quality spectra are
well consistent with the macronova/kilonova model (initially it was dominated by the lanthanide-free outflow region
that may be mainly contributed by the accretion disk wind or the neutrino-driven mass loss of the hypermassive neutron star formed in
the merger; and at late times it was dominated by the emission from the lanthanide-rich region). To the surprise of the community, a remarkable GW/GRB/macronova association
is firmly established in the first GW event involving neutron star(s). The long-standing prediction that neutron star mergers are the sources of
short duration GRBs \citep{1989Natur.340..126E} has thus been directly confirmed. Moreover, the GW/GRB/macronova association has some far-reaching implications
for both physics and astrophysics, which are the focus of this work.

After the claim of the possible detection of a transient associated with GW150914 by Fermi-GBM
\citep{2016ApJ...826L...6C}, we had discussed some
implications of the transient/GW association \citep{2016ApJ...827L..16L}. This work extends our previous
approaches significantly. In addition to comparing GRB 170817A to other sGRBs and measure the velocity of the GW,
we further test the Einstein Equal Principle (i.e., for the specific scenario that the
photons and GWs may not follow the same trajectories in the gravitational field), and rule out ``the Dark Matter
Emulators and some dark energy models". Moreover, with the unambiguous detection of a large amount of the r-process
elements in the macronova associated with GW170817A, we show that the neutron star mergers are indeed the main sites of
the very heavy elements in the Universe and the binary strange star merger model for GRB 170817A is ruled out.

\section{GRB170817A and the previous SGRBs}\label{sec:compare}

\citet{2016ApJ...827L..16L} suggested to test the merger origin of old sGRBs via the comparison with
the newly detected GRBs/GW events.
If these GW-associated GRB events  are
found to be similar to the (old) events without GW observation data in
many aspects, the merger scenario for sGRBs
may be supported. Though such a test is likely non-trivial,
one of the cautions is that the advanced LIGO/Virgo can only reach $z\leq 0.1$ for neutron star mergers.
For such local events, some merger-driven GRBs can be detectable even when
our line of sight is outside the cone of the ``uniform" relativistic ejecta
or a bit far from the symmetric axis of the structured outflow \citep[e.g.][]{2017arXiv170807008J,2017arXiv170807488K,Yamazaki2002}.
The shock breakout of relativistic ejecta from surrounding sub-relativistic outflow launched during the
merger may also generate some under-luminous GRBs \citep[e.g.][]{Kasliwal2017}.
Therefore, the GW-associated GRBs are likely dominated by an apparently ``under-luminous" group
and the goal outlined in \citet{2016ApJ...827L..16L} may be potentially achievable only when a sub-group of bright local sGRBs
have been detected.

\begin{figure}[ht!]
\figurenum{1}\label{fig:relation}
\centering
\includegraphics[angle=0,scale=0.42]{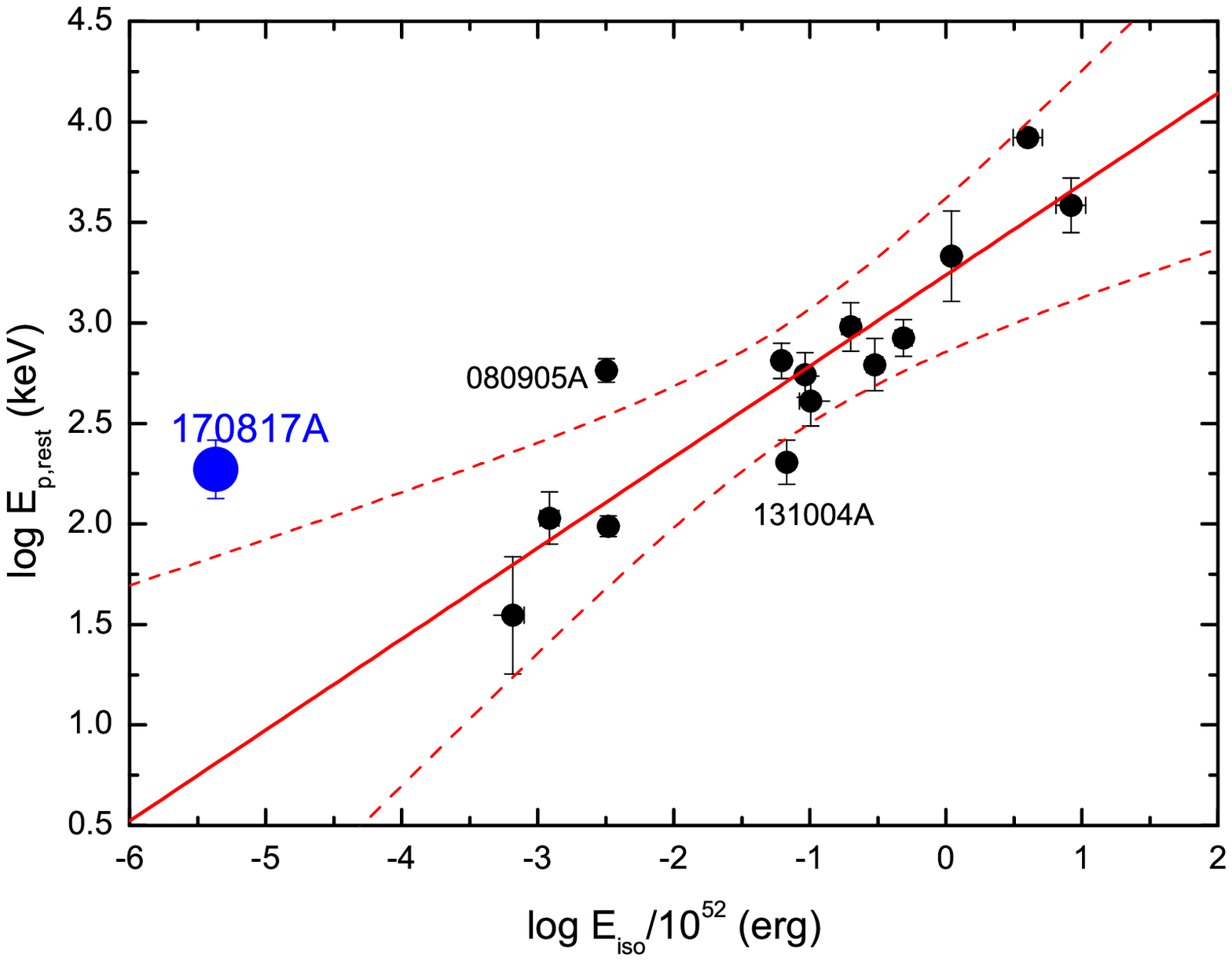}
\includegraphics[angle=0,scale=0.42]{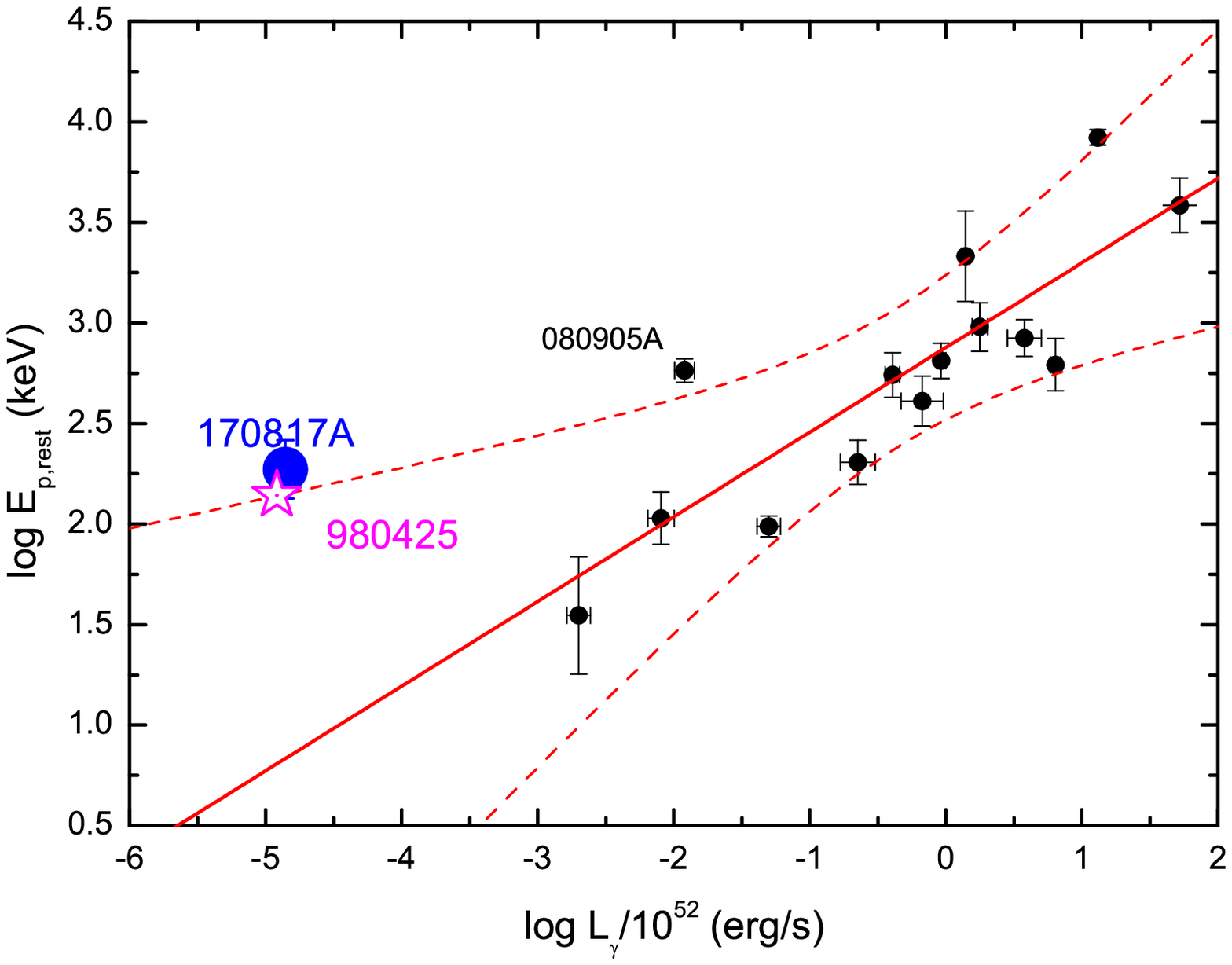}
\caption{The left and right panels present the correlations between the rest frame spectral
peak energy $E_{\rm p,rest}$ and the isotropic total energy $E_{\rm iso}$ and the luminosity
$L_\gamma$ of SGRBs, respectively.
The solid lines are the best-fit correlations, i.e., $\log E_{\rm p,rest}=(3.24\pm0.10)+(0.45\pm0.06) \log(E_{\rm iso}/10^{52}{\rm erg})$ and $\log E_{\rm p,rest}=(2.88\pm0.10)+(0.42\pm0.08) \log(L_\gamma/10^{52}{\rm erg~s^{-1}})$, while the dashed lines represent  3-sigma scatters. Only the SGRBs with well measured spectra are included.
Clearly GRB 170817A does not well follow these two correlations. The data of GRB 980425 and GRB 170817A are adopted from \citet{Ghisellini2006} and \citet{2017ApJ...848L..14G}, respectively. Other data are either taken
from \citet{ZhangFW2012} and \citet{Gruber2014} or analyzed in this work.}
\hfill
\end{figure}

Since GRB 170817A is the first short burst unambiguously associated with a GW event, it is necessary to be compared with other sGRBs.
Following \citet{2016ApJ...827L..16L} we present the $E_{\rm p,rest}-E_{\rm iso}$ and $E_{\rm p,rest}-L_\gamma$ diagrams, where $(E_{\rm p},~E_{\rm iso},~L_\gamma)$ are the (spectral peak energy, isotropic equivalent energy, luminosity) of the prompt emission, respectively, and the subscript ${\rm rest}$ represents the parameter(s) measured in the host galaxy frame of the burster. Only the sGRBs with the well measured spectra are included.
As shown in the Fig.\ref{fig:relation}, GRB 170817A is the weakest sGRB detected so far and its $E_{\rm iso}$ and $L_\gamma$ are more than two orders of magnitude lower than those recorded before. However, its $E_{\rm p,rest}=187 \pm 63$ keV \citep{2017ApJ...848L..14G} is comparable to quite a few sGRBs \citep{2017ApJ...848L..14G}. Therefore,  GRB 170817A do not follow
the regular correlations (see the solid lines in Fig.\ref{fig:relation}). One possible interpretation is that GRB 170817A is an off-beam/off-axis event or a shock breakout event.
In Fig.\ref{fig:relation} we have also compared sGRB 170817A and long event GRB 980425, the closest bursts in each group. Surprisingly, sGRB 170817A and GRB 980425, two events with completely different progenitors, have rather similar $L_\gamma$ and $E_{\rm p,rest}$ (see Fig.\ref{fig:relation}). If not just a coincidence, this might indicate similar radiation processes. The progenitor of GRB 980425 is known to be a massive star. It's prompt radiation process is still unclear and an attractive model is the shock breakout of relativistic outflow from the stellar envelope with a significant density gradient \citep{Kulkarni1998}. For sGRB 170817A originated from a neutron star binary merger, there was certainly no stellar envelope. The numerical simulation suggest that the sub-relativistic outflow launched during the merger can play a similar role and GRB 170817A could be a shock breakout event \citep{Kasliwal2017}.

\section{Time lag between the GW and GRB signals: astrophysical and physical implications}\label{sec:lag}

\subsection{The astrophysical implications of the $\sim 1.7$ s time lag between the GW and GRB signals}\label{subsection:time lag}
In \citet[][Sec.3 therein]{2016ApJ...827...75L} the model-dependent time delay between the GW and GRB signals (i.e., $\Delta t_{\rm GW-GRB}$) has been extensively investigated. As summarized in their Tab.1,
the general prediction is $\Delta t_{\rm GW-GRB} \sim 0.01-{\rm a~few}$ seconds, depending on the collapse time of the hypermassive/supramassive remnant formed in the binary neutron star mergers and on the energy dissipation process/radius.
The $\sim 1.7$ s time delay between GW170817 and GRB 170817A is in agreement with the previous predictions. It could indicate the thermal support scenario that the hypermassive/supermassive neutron star did not collapse until the neutrinos have leaked out in a timescale of $\sim 1$s, or the (magnetic) energy dissipation took place at $\sim 10^{15}-10^{16}$ cm, or our line of sight is away from the ejecta edge (the angle is $\Delta \theta$) and the prompt emission started at a radius of $\sim 4.5\times 10^{12}~{\rm cm}~(\Delta \theta/0.15)^{-2}$.

At least for GW170817/GRB 170817A, the specific model developed to explain the sGRBs with extended X-ray emission, which predicts $\Delta t_{\rm GW-GRB}\sim 10^{2}-10^{4}$ s \citep{2015MNRAS.448.2624C, 2015ApJ...802...95R}, has been ruled out. With a reasonably large GW/GRB association sample expected in the next decade, it will be extremely interesting to see whether the distribution of $\Delta t_{\rm GW-GRB,int}$ is narrow or wide, or even highly structured, with which the long $\Delta t_{\rm GW-GRB}$ model can be partly confirmed or unconvincingly ruled out.

\subsection{Measuring the GW velocity, testing the equivalence principle and ruling out some modified gravity models for dark matter and dark energy}\label{subsec:vg}
\subsubsection{Measuring the GW velocity}\label{subsubsec:velocity}
In some modified gravity theories amazing to explain away dark matter or dark energy, GW travels in the vacuum at velocities that can be different from the speed of light \citep[i.e., $\varsigma \equiv (c-v_g)/c\neq 0$; see e.g.][for reviews]{2012PhR...513....1C,Joyce2015}.
In this work we assume a constant $\varsigma$.
The sub-luminal movement of gravitons (i.e., $\varsigma>0$) has already been tightly constrained by the absence of gravitational Cerenkov radiation of ultra-high energy cosmic rays \citep{2001JHEP...09..023M}
%(i.e., $0\leqslant \varsigma \leqslant 2 \times 10^{-15}$ for galactic origin; $0\leqslant \varsigma \leqslant 2 \times 10^{-19}$ for extragalactic origin; see \cite{2001JHEP...09..023M})
\footnote{In order to ``save" some dark energy models, it is argued in some literature that currently no extra-galactic source of ultra-high energy cosmic rays has  been identified yet and these particles may have a galactic origin, for which the Vainshtein screening mechanism is at play and the above constraint can not be applied to the cosmological data \citep{2017A&A...600A..40N}. However, the GW170817/macronova association sets an independent stringent constraint on the sub-luminal movement of gravitons (see eq.(\ref{eq:constr-2})).\label{footnote-1}}.
The superluminal constraints of gravitons (i.e., $\varsigma<0$) are weak and model-dependent \citep{2014PhRvD..89h4067Y,2016PhRvL.116f1101B,2015JCAP...03..016A,2016JCAP...02..053B}. %Model-independent but very weak constraint can be derived from the time delay of the signals between the GW detectors \citep{2016JETPL.103..624B}.
The simultaneously radiated GW and electromagnetic signals can set stringent/robust constraint on $\varsigma$ \citep{1998PhRvD..57.2061W,2014PhRvD..90d4048N,2016ApJ...827...75L} because after traveling a distance of $D\sim 10^{2}$ Mpc, even a very tiny $ \varsigma$ will induce a time delay of
$\Delta t_{\varsigma} \approx 1~{\rm s}~ ({\varsigma \over 10^{-16}}) ({D \over 100~\rm Mpc})$.
Note that in the absence of equivalence principle violation, $\Delta t_{\rm GW-GRB}=\Delta t_{\varsigma}+\Delta t_{\rm e}$, where $\Delta t_{\rm e}$ represents the intrinsic delay of the emitting times of the GW signal and the GRB.
In the merger-driven scenario, the GW single always precedes the GRB emission and we have $\Delta t_{\rm e}\geq 0$ and hence $\Delta t_{\rm GW-GRB}\geq \Delta t_{\varsigma}$.
For GW170817/GRB170817A with $\Delta t_{\rm GW-GRB}\sim 1.7$ s and $D\sim 40$ Mpc, the constraint reads \citep[see also][]{2017ApJ...848L..13A}
\begin{equation}
-4.3\times 10^{-16}\leq \varsigma\leq 0.
\label{eq:constr-1}
\end{equation}
Such results imply that the super-luminal movement of gravitons, if any, should not exceed the speed of light by a velocity of $1.3\times 10^{-5}~{\rm cm~s^{-1}}$.

\begin{figure}[ht!]
\figurenum{2}\label{fig:varsigma}
\centering
\includegraphics[angle=0,scale=0.5]{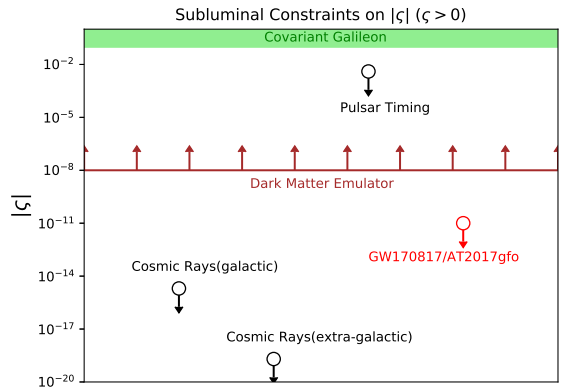}
\caption{Constraints on the sub-luminal movement of the gravitational wave. The cosmic ray constraints are adopted from \citet{2001JHEP...09..023M}, where the weaker constraint refers to the Galactic origin model and the much stronger constraint is for the extragalactic origin model of the ultra-high energy cosmic rays. The pulsar timing constraint is taken from \citet{2008PhRvD..78d4018B}. Interesting, the Dark Matter
Emulators and a class of dark energy models (``Covariant Galileon") have been ruled out at high confidence levels (see Section \ref{subsec:rule out}).}
\hfill
\end{figure}

A reliable constraint on the sub-luminal movement of GWs with $\Delta t_{\rm GW-GRB}$ for a single GW/GRB association event is less straightforward. This is because $\Delta t_{\rm e}$ could be long (for instance $\sim 10^{2}-10^{4}$ s or even longer, as speculated in \cite{2015ApJ...802...95R}), which hampers a reliable constraint on $\varsigma$. The problem can be solved in the (near) future when NS-BH merger driven GW/GRB event has been successfully detected, for which a small $\Delta t_{\rm e}<T_{90}$ is predicted, where $T_{90}$ is the duration of the prompt emission of the GRB \citep{2016ApJ...827...75L}. In the current case, the constraint on the sub-luminal movement of GWs is still possible since the optical emission of macronova/kilonova is known to present within 1 day after the merger \citep{2013ApJ...774...25K,2013ApJ...775...18B}. The successful detection of macronova/kilonova emission at $t_{\rm mn,det}\sim 0.5$ days suggest that the time delay of the arrival of the GW signal due to its sub-luminal movement can not be longer than $\sim 0.5$ days, then we have
\begin{equation}
0\leq \varsigma \leq 10^{-11}(t_{\rm mn,det}/4\times 10^{4}~{\rm s}).
\label{eq:constr-2}
\end{equation}
\citet{2017ApJ...848L..13A} reported a much tighter constraint on the sub-luminal movement of gravitons by (arbitrarily) assuming a $\sim 10$s intrinsic delay between the merger and the prompt GRB emission. Our constraint is weaker but less assumption-dependent.
%Please note that if the sub-luminal movement of gravitons is due to a non-zero mass,
%very stringent constraint can be set by the almost simultaneous arrival times of the GWs at different frequencies \citep[e.g.][]{2016PhRvL.116f1102A}.
In Fig. \ref{fig:varsigma} we show our bound in comparison to some previous constraints.

\subsubsection{Testing the Einstein equivalence principle}\label{subsubsec:eep}
Another scenario yielding the different arrival time of ``simultaneous" emitted GWs and photons, two different types of massless particles, is the violation of Einstein equivalence principle (EEP). In the framework of parameterized post-Newtonian approximation, deviations from EEP can be described by a parameter $\gamma$, which is 1 in general relativity. Therefore, the GW/GRB association is very suitable to test the EEP violation \citep{1999BASI...27..627S,2016PhRvD..94b4061W,2016ApJ...827...75L}. The Shapiro delay is generally calculated as
$\Delta t_{\rm gra}=-\frac{\Delta \gamma}{c^3}\int_{r_{\rm o}}^{r_{\rm e}}U(r(t); t)$ \citep{1964PhRvL..13..789S,1988PhRvL..60Q.176K,1988PhRvL..60..173L}, where the integral is along the travelling path of photons and $U(r(t); t)$ is the gravitational potential.
The time delay caused by milky way can be calculated by $\Delta t_{\rm gra}=1.7 \times 10^{7}~{\rm s}~ \Delta\gamma  ( {M_{\rm MW}}/{6\times 10^{11} M_{\odot}}) (\log(D/b)/{4\log10})$ \citep{1973grav.book.....M,1988PhRvL..60..173L},
where $\Delta \gamma \equiv \gamma_{\rm photon}-\gamma_{\rm GW}$ (if $\Delta \gamma\neq 0$, it would mean that the photons and GWs do not follow the same trajectories in the gravitational field of the galaxy and the EEP is violated), $M_{\rm MW}$ is the total mass of milky way and $b$ is the impact parameter of the particle paths relative to the center of the milky way.
Now the observed $\Delta t_{\rm GW-GRB}$ should be expressed as
$\Delta t_{\rm GW-GRB}=\Delta t_{\rm e}-\Delta t_{\varsigma} + \Delta t_{\rm gra}$.
We thus need a group of GW/GRB events, in particular those driven by NS-BH mergers, at different $D$ to self-consistently constrain $\varsigma$ and $\Delta \gamma$. This is because
for NS-BH merger driven GW/GRB events it is generally expected that $\Delta t_{\rm e}\leq T_{90}$ \citep[please see][for the extensive discussion]{2016ApJ...827...75L}. For the current data and under the assumptions of $\varsigma=0$ (i.e., in the vacuum the GW velocity equals to the speed of light) and $\Delta t_{\rm e}=0$ (i.e., $\Delta t_{\rm GW-GRB}=\Delta t_{\rm gra}$), a rough constraint on $\Delta \gamma$ reads
\begin{equation}
\Delta \gamma \leq 10^{-7}~({\Delta t_{\rm GW-GRB}\over 1.7~{\rm s}})\left(\frac{M_{MW}}{6\times 10^{11} M_{\odot}}\right)^{-1} [\frac{\log(D/b)}{4\log10}]^{-1}.
\end{equation}
Such a constraint can be further improved. As noticed in \citet{Nusser2016}, the potential
fluctuations from the large scale structure, which can be found from the observed peculiar
velocities (deviations from a pure Hubble
flow; $v_{\rm p}$) of galaxies, are significantly larger than the gravitational
potential of the Milky Way ($U_{\rm MW}$). Peculiar velocity data yield a bulk peculiar velocity of
$v_{\rm p}\sim 300~{\rm km~s^{-1}}$ for the sphere of radius $R\sim 50$ Mpc
around us \citep{MaYZ2013}, suggesting a gravitational potential
$U\sim v_{\rm p} R H_{0} \sim 25U_{\rm MW}$ at the site of GW170817/GRB170817A, where $H_{0}$ is the Hubble's constant.
Therefore, for the current data we have a constraint
\begin{equation}
\Delta \gamma \leq 4\times 10^{-9}.
\end{equation}
Such a constraint has taken into account the contribution of the gravitational potential of the large scale structure, which is thus stronger than the bound inferred from the better measured Milky Way gravitational potential alone \citep[see also][]{2017ApJ...848L..13A,2017JCAP...11..035W}. 

Here we simply adopt the GW/GRB association to set the bound. In the future, {\it if the strong gravitational lensing of GW/GRB association events can be detected as well,
one can use the delay times of the GW/GRB signals and their corresponding lensing ``counterparts" to set stringent constraint on $\Delta \gamma$.} This is because the gravitational wave potential of the lens (in particular the galaxy clusters) will induce an additional Shapiro delay if $\Delta \gamma \neq 0$. The main challenge for such an approach is however the absence/rarity of such events in the foreseeable future.

\subsubsection{Ruling out Dark Matter
Emulators and some dark energy models}\label{subsec:rule out}
In general relativity, the GW velocity is the same as the speed of light.
However, major outstanding theoretical issues such
as the nature of dark energy and dark matter have led
to consider the possibility that gravity differs from GR
in some regimes \citep[see][for reviews]{2012PhR...513....1C,Joyce2015}.
Some of these models predict very different arrival times of the simultaneously radiated GW/GRB signals and hence can be accurately tested.

For example, motivated by the non-detection of dark matter particles so far, there are a group of modified gravity theories, known as dark matter emulators, which dispense with the need for dark matter. These models have the property that weak GWs couple to the metric that would follow from general relativity without dark matter whereas ordinary particles couple to a combination of the metric and other fields which reproduces the result of general relativity with dark matter. The absence of reliable detection of dark matter particles so far renders such a possibility attractive. \cite{2008PhRvD..77l4041D} show that there is an appreciable difference in the Shapiro delays of GWs and photons from the same source, with the GWs always arriving first.
Even for the very nearby extragalactic sources, the predicted time-lags between the GW signals and the electromagnetic counterparts ($\Delta t_{\rm DME}$) are several hundreds of days.  Additional comparable time-lag arises during the propagation in the host galaxy of the source. If it is indeed the case, in the extragalactic space the GW should move subluminally to yield an almost simultaneous arrival of GW170817 and GRB 170817A, i.e., $\Delta t_{\varsigma}+\Delta t_{\rm DME}\approx 0$, which then yields
\begin{equation}
\varsigma \sim 2.1\times 10^{-8}({D\over 40~{\rm Mpc}})({\Delta t_{\rm DME}\over 10^{3}~{\rm days}}).
\end{equation}
Such a $\varsigma$, however, is already about 3 orders of magnitude larger than our subluminal bound set by GW170817/AT2017gfo in eq.(\ref{eq:constr-2}). The tension is far stronger (i.e., the divergency is by a factor of $\sim 10^{7}$ or more) if the submuminal movement constraints set by ultrahigh-energy cosmic rays applies (see however footnote \ref{footnote-1}).
We therefore conclude that the Dark Matter Emulators has been ruled out and the dark matter model is favored (See Fig.\ref{fig:varsigma}).

There is also a large class of scalar-tensor theories, which predict that GWs propagate with velocity different from the speed of light and a difference of ${\cal O}(1)$ is possible for many models of dark energy. For example, in the model of ``Covariant Galileon"\citep{2009PhRvD..79h4003D,2014JCAP...08..059B}, the violation parameter is about $\varsigma \sim 10\% - 100\%$, and the delay between GW and electromagnetic signals from distant events will run far beyond human time-scales \citep{2016JCAP...03..031L,2017PhLB..765..382L,2017PhRvD..95h4029B}, which is clearly not the case for GW170817/GRB 170817A. The first GW/GRB association event thus places very stringent
constraint on theories allowing variations in the speed of
GWs and eliminates many contender models for cosmic acceleration (See Fig.\ref{fig:varsigma}).

\section{Implications for the r-process element origin and Constraining the double strange star merger model}\label{sec:macronova}

\subsection{Neutron star mergers as the main site of the r-process element production}\label{subsection:r-process}
\begin{figure}[ht!]
\figurenum{3}\label{fig:heavy elements}
\centering
\includegraphics[angle=0,scale=0.5]{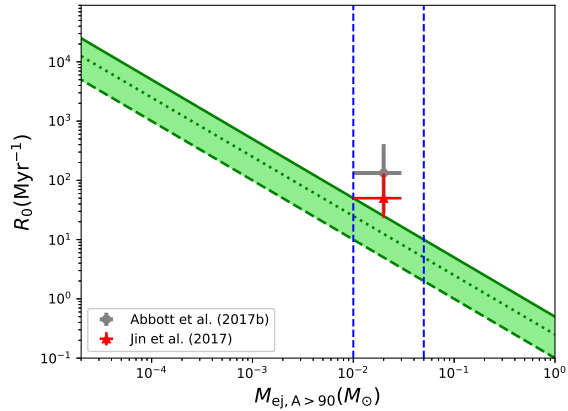}
\caption{The binary neutron star merger rate and ejected mass inferred from current GRB/macronova observations, in comparison to what needed to reproduce the Milky Way r-process material. The green solid, dotted, dashed lines are $R_0=(1.0,~0.5,~0.2)\langle R \rangle$, respectively. The blue vertical dotted lines represent the ejecta mass range inferred from current macronova modeling, assuming that $\sim 1/2$ of the ejected material with $A\geq 90$. The data points represent the neutron star merger rates and heavy element mass of GW170817. If a  $\sim 0.04M_\odot$ ejecta mass is typical for the mergers, the rate in local universe is likely higher than that ``needed" in the galaxy, which may imply a  merger rate of our Milky Way lower than other galaxies or typically $M_{\rm ej}\sim 0.01M_\odot$.}
\hfill
\end{figure}

The heavy elements origin, also known as nucleosynthesis, is one of the mysteries in the universe \citep{2007PhR...442..237Q}. The widely discussed sites include the core collapse supernovae \citep{1957RvMP...29..547B} and neutron star mergers \citep{1974ApJ...192L.145L,1989Natur.340..126E}. Though there are increasing evidence that the neutron star mergers are significant site of the heavy elements \citep[e.g.][]{2013Natur.500..547T,2015NatCo...6E7323Y,2015NatPh..11.1042H,2016Nat...531...610,2016NatCo...712898J}, the unambiguous detection of a large amount of r-process material in AT2017gfo provides the most direct evidence \citep{Pian2017,2017Natur.551...80K,2017arXiv171005443D,2017ApJ...848L..19C}. To account for the measured total mass of galactic heavy r-process elements (i.e. $A>90$), the binary neutron star merger rate averaged over the galaxy age should be $\langle R\rangle \approx 50~\mathrm{Myr^{-1}}~({M_{\rm ej,A>90}}/{0.01M_{\odot}})^{-1}$ \citep{2015NatPh..11.1042H}, where $M_{\rm ej, A>90}$ refers to the heavy element mass ejection for a single event. However, the merger rate is actually not a constant and the inferred merger rate at present time ($R_{\rm 0}$) may be lower than the averaged one by a factor of a few (i.e., $R_{\rm o}<\langle R\rangle$). Thus we draw the lines of $R_0=(1, ~0.5,~0.2)\langle R\rangle$ in Fig.\ref{fig:heavy elements}.

Based on four ``nearby" sGRBs with reasonably estimated jet half-opening angles, a {\it conservative} estimate of the local ($z\leq 0.2$) neutron star merger rate is $\sim 583^{+923}_{-318}~{\rm Gpc^{-3}~yr^{-1}}$ \citep{2017arXiv170807008J}. Such a (conservative) merger rate is well consistent with that ($\sim 1540^{+3200}_{-1220}~{\rm Gpc^{-3}~yr^{-1}}$) inferred from the successful detection of a neutron star merger event by advanced LIGO/Virgo in their second observational run \citep{LVC2017}.
Since the density of the Milky Way Equivalence Galaxy density in the local universe is $\sim 1.16 \times 10^{-2}~\rm Mpc^{-3}$ \citep{2008ApJ...675.1459K}, we can convert the number to the Milky Way merger rate $\sim 50^{+80}_{-27} \rm Myr^{-1}$. On the other hand, the macronova spectrum modeling suggests the mass ejection of GRB170817 to be $M_{\rm ej}\sim 0.04\pm0.01~M_\odot$ and the heavy r-process material may consist of $\sim 1/2$ of the total \citep{Pian2017}. These information have been presented in Fig.\ref{fig:heavy elements}.  The ``data point" is above the line of $R_0=\langle R\rangle$, which is in support of the neutron star merger origin of the r-process material \citep[see also e.g.][]{2017Natur.551...80K,2017arXiv171005443D,2017ApJ...848L..19C} and furthermore implies that either the ``averaged" rate of neutron star mergers in the Milky Way is lower than that of the ``local" Universe or the ``typical" mass ejection of such mergers is significantly smaller than $0.04M_{\odot}$. For the latter, one caveat is that the neutron-rich outflow mass estimated for GRB 130603B, GRB 060614 and GRB 050709 are in the range of $\sim 0.02-0.1M_\odot$ \citep{2013Natur.500..547T,2015NatCo...6E7323Y,2016NatCo...712898J}.

\subsection{Constraining the double strange quark star merger model for GRB 170817A} \label{subsec:strange}

Strange matter made of quarks may be the ground state of matter, and neutron stars with the sufficiently-high central density may become strange stars \citep{1970PThPh..44..291I,1986ApJ...310..261A}. If strange stars exist, there may be some binary strange star systems. Thanks to the orbital decay resulting in the energy and angular momentum loss via GW, some systems will merger within the Hubble timescale and give rise to GW events and short duration gamma-ray bursts \citep{1991ApJ...375..209H,2017arXiv171004964L}. During the merger phase of binaries stars, some strange matter are injected into the interstellar medium. The mass distribution outcome of the fragmentation of the strange matter has been investigated \citep{2014PhLB..733..164P} and the expected nucleosynthesis spectra for the strange star-strange star merger scenario have been calculated \citep{2017IJMPS..4560042P}. Different from the neutron star mergers, no significant r-process nucleosynthesis is expected since the high temperature deconfinement of strange matter would produce large amounts of neutrons and protons and the mass buildup would proceed in a Big-Bang nucleosynthesis like scenario. In particular, the neutron to proton ratio (typically $\sim 0.7$) would allow to reach the iron peak only \citep{2017IJMPS..4560042P}. The decay of the heavy elements will still heat the outflow and yield optical transient. The absence of lanthanides however does not result in a relatively long-lasting infrared bump. Moreover, the spectrum should be significantly different from the neutron star merger-driven kilonova/macronova.
The high quality kilonova/macronova spectra, in particular those measured at late times \citep{Pian2017}, however are well consistent with the synthetic spectra of r-process material model \citep{2013ApJ...774...25K}. The double strange star merger scenario for GW170817/GRB 170817A is thus convincingly ruled out.

\section{Summary} \label{sec:discussion}
The GW/GRB/macronova association established in August 2017 directly confirms the long-standing suggestions that neutron star merges do take place frequently and generate strong GWs, which further produce short gamma-ray flashes and launch r-process material. In this work we have discussed some far-reaching additional physical and astrophysical implications. In particular, we show that:
\begin{itemize}
\item The short time delay between the GW and GRB signals set a very tight constraint on the possible superluminal movement of GWs and the difference between its velocity and the speed of light should be within a factor of $\sim 4.3\times 10^{-16}$ \citep[see also][]{2017ApJ...848L..13A}. The GW/macronova association set an independent constraint on the possible subluminal movement of GWs and the difference between its velocity and the speed of light should be within a factor of $\sim 10^{-11}$. The underlying assumption for these constraints is that the GW velocity is independent of the frequency (see Section \ref{subsubsec:velocity}). In the foreseeable future, these two constraints can be improved by (quite) a few orders of magnitude.
\item The possible violation of weak equivalence principle is tightly constrained (the additional assumption is that in the vacuum the GW velocity equals to the speed of light) and the difference of the gamma-ray and GW trajectories in the gravitational field of the galaxy and the local universe should be within a factor of $\sim 3.4\times 10^{-9}$ (see Section \ref{subsubsec:eep}).
\item The so-called Dark Matter Emulators and some contender models for cosmic acceleration, such as ``Covariant Galileon", which predicted long time delay of the arrival times of the simultaneously radiated GWs and photons from the same source, are ruled out (see Section \ref{subsec:rule out} and Fig.\ref{fig:varsigma}).
\item The high neutron star merger rate (inferred from both the local sGRB data and the GW data) together with the significant ejected mass strongly suggest that such mergers are main sites of heavy r-process nucleosynthesis (see Section \ref{subsection:r-process} and Fig.\ref{fig:heavy elements} and also  \citet{2017Natur.551...80K}, \citet{2017arXiv171005443D}, \cite{2017ApJ...848L..19C}). {Moreover, it is likely that the ``averaged" rate of neutron star mergers in the Milky Way is lower than that of the ``local" Universe.}
\item The successful identification of Lanthanide elements in the macronova/kilonova spectrum also excludes the possibility that the progenitors of GRB 170817A are a binary strange quark star system (see Section \ref{subsec:strange}).
\end{itemize}

Finally, we'd like to mention the puzzling fact that sGRB 170817A ($D\sim 40$ Mpc) and GRB 980425 ($D\sim 36$ Mpc), two events with completely different progenitors, have almost the same $L_\gamma$ and $E_{\rm p,rest}$ (see Fig.\ref{fig:relation}), which might indicate similar prompt radiation processes if not just a coincidence.

\section*{Acknowledgments}
We thank Yi-Ming Hu and Yi-Fan Wang for useful discussions and the anonymous referee for helpful suggestions.
This work was supported in part by 973 Programme of China (No. 2014CB845800), by NSFC under grants 11525313 (the National Natural Fund for Distinguished Young Scholars), 11273063 and 11433009, by the Chinese Academy of Sciences via the Strategic Priority Research Program (No. XDB09000000) and the External Cooperation Program of BIC (No. 114332KYSB20160007).

\clearpage

\end{document}